\documentclass[12pt,epsfig,eps]{article}
\usepackage{epsfig}
\usepackage{graphicx}
\usepackage{amssymb}
\usepackage{amsmath}
\def\permil{\%\raise.10ex\hbox{$_{\scriptstyle 0}$}}

\def\beq{\begin{equation}}
\def\eeq{\end{equation}}
\def\beqn{\begin{eqnarray}}
\def\eeqn{\end{eqnarray}}
\newcommand{\mbf}[1]{\mbox{\boldmath $#1$}}

\newcommand{\bk}{\mbf{k}}

\newcommand{\bl}{\mbf{l}}
\newcommand{\bq}{\mbf{q}}
\newcommand{\bp}{\mbf{p}}

\def\bea{\begin{eqnarray}}
\def\eea{\end{eqnarray}}
\begin{document}
\hfill
\hspace*{\fill}
\begin{minipage}[t]{3cm}
  DESY-11-074
\end{minipage}
\vspace*{1.cm}
\begin{center}
\begin{Huge}
Recombination within multi-chain contributions in $pp$ scattering\\[1cm]
\end{Huge}
\begin{large}
\vspace{0.5cm}
J. Bartels$^a$ and M.G.Ryskin$^b$\\[1cm] 
$^a$ II. Institut f\"{u}r Theoretische Physik, Universit\"{a}t Hamburg,
Luruper Chaussee 149, D-22761 Hamburg, Germany\\
$^b$ Petersburg Nuclear Physics Institute, Gatchina, St.Petersburg,
188300, Russia
\end{large}
\end{center}
\vskip15.0pt \centerline{\bf Abstract}
We investigate the evolution of multiple parton chains in proton-proton scattering and
show that interactions between different chains may become quite important. 
%%%%%%%%%%%%%%%%%%%%%%%%%%%%%%%%%%%%%%%%%%%%%%%%%%%%%%%%%%%%%%%%%%%%%%%%%%%%
\section{Introduction}
In the last years it has become clear that multiple parton interactions 
play an important role in hadron-hadron collisions 
at high energies \cite{Treleani:2007gi,Berger:2009cm, Gaunt:2009re,
Gaunt:2010pi, Strikman:2010bg, Diehl:2011tt, Flensburg:2011kj,Ryskin:2011kk}.
As a first step these 
chains are modelled as a collection of single noninteracting chains (Fig.1).
\begin{figure}[t]
\begin{center}
\epsfig{file=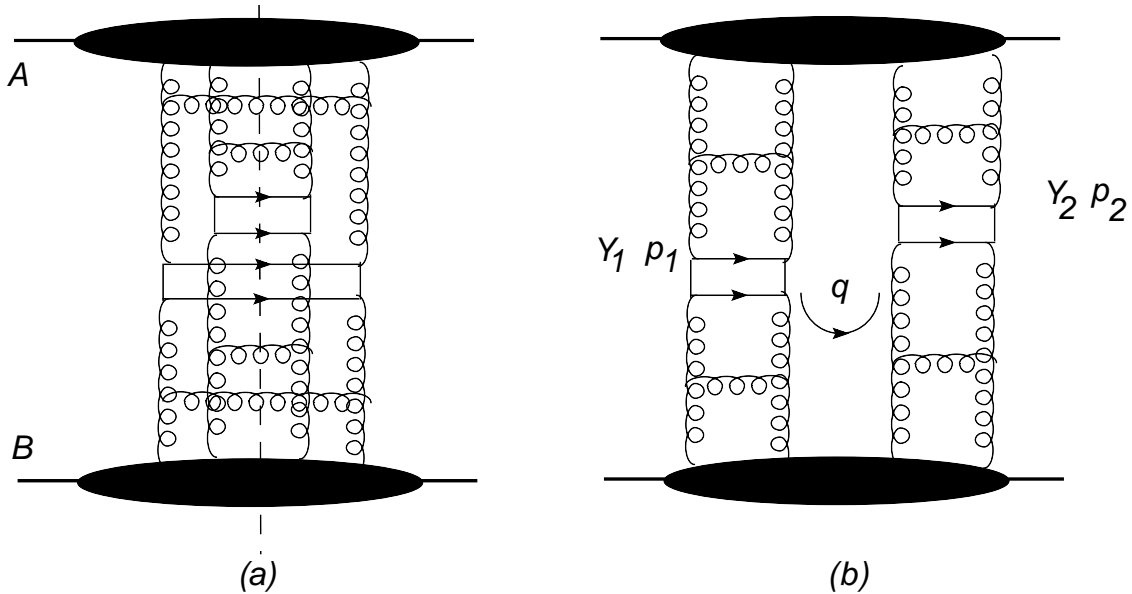,width=15cm,height=8cm} 
\caption{Two noninteracting chains: (a) the cross section (energy discontinuity
of the scattering amplitude $T_{2 \to 2}$), (b) redrawn as a (uncut) 
two-ladder exchange diagram.   
Within each chain, the boxes mark the hard subprocess with largest 
transverse momenta (production of dijets).}
\end{center}
\end{figure}
Each chain follows the usual partonic DGLAP evolution, i.e. the ladders 
are in color singlet states, and the momentum transfer across the ladder 
is set equal to zero. Disregarding final state radiation and working in 
leading order only, the final state produced by $k$ such chains consists of 
$N=n_1+n_2+...+n_k$ partons, and the cross section, $d\sigma \sim |T_{2 \to  N}|^2 d\Omega_N$,  
is described by a sum of squares, without any interference terms.

The theory of multiparton (higher twist) evolution has been outlined in 
\cite{BFKL}. To leading order, the evolution is described by the sum over 
the pairwise interactions between two $t$-channel partons, and the evolution kernels 
are given by the nonforward DGLAP splitting functions. Of particular 
importance is the small-$x$ region where powers of $\ln 1/x$ may compensate 
(and even overcome) the 
higher twist suppression. In this region, the dominant contributions are 
given by gluon ladders, and their evolution in $\ln 1/x$ is described by 
the BKP equations \cite{BKP}. In leading order this evolution is, again, 
described by the sum over parwise interactions between $t$-channel gluons.
The evolution kernels are given by the nonforward BFKL-kernels.
At small $x$, the leading logarithmic approximations of the two approaches 
- higher twist evolution in momentum scale or small-$x$ evolution in $\ln 1/x$ - 
coincide in the so-called 'double logarithmic approximation' which samples 
powers of $(\alpha_s \ln 1/x  \ln p^2)$. 
         
For both evolution schemes the t-channel multiparton state is in a color 
singlet state and, as far as the total cross section is concerned, 
the total momentum transfer is set equal to zero. However, any 
subsystem consisting of two $t$-channel gluons, in general, will have nonzero color 
quantum number and nonzero momentum transfer. Therefore, describing the 
evolution of a $t$-channel state consisting of, say, $2n$ gluons 
as the evolution of $n$ noninteracting color singlet ladders with zero momentum transfer represents an approximation 
whose validity deserves further investigation.
  
In this note we study, as a first correction beyond the approximation 
of noninteracting ladders, a particular 
type of 'interactions' between two ladders which we illustrate in Fig.2:  
\begin{figure}[t]
\begin{center}
\epsfig{file=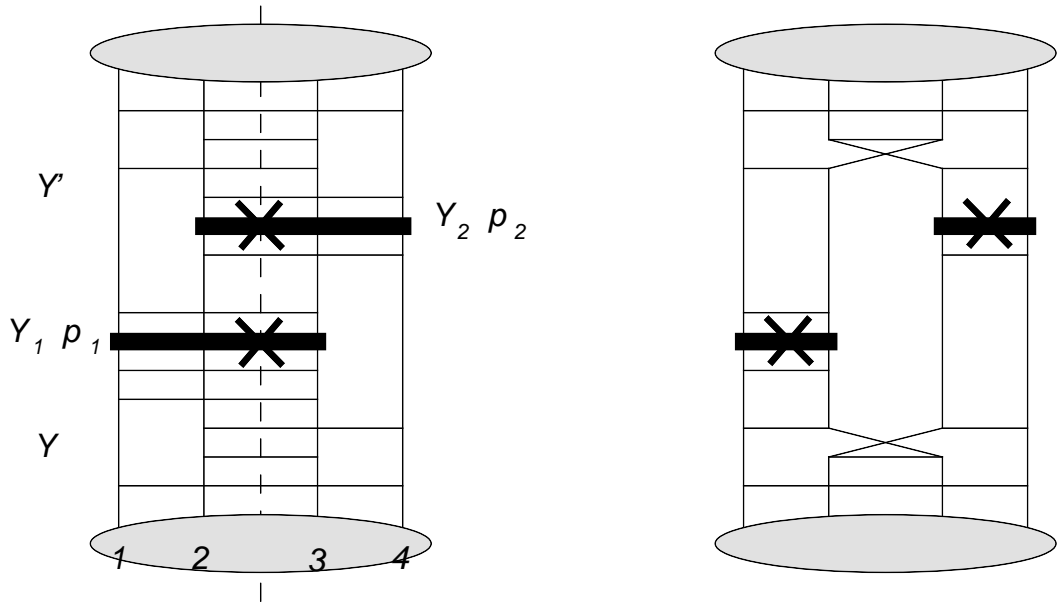,width=14cm,height=8cm}\\ 
\caption{Two recombinations within the two chains.} 
\end{center}
\end{figure}
Starting at the proton at the bottom of Fig.2, we first have two noninteracting color 
singlet ladders (denoted by the pairs of $t$-channel gluons $(14)$ and $(23)$). 
At rapidity $Y$ we allow for a 'recombination' of $t$-channel gluons: 
from now on we have the two color singlet pairs  $(13)$ and $(24)$. 
In the following we will denote his transition by 'recombination vertex'. 
It introduces a correlation between the two ladders. However, it is important 
to note that, in the double log approximation, this kind of interaction 
between the two ladders still belongs to the leading logarithmic approximation: 
for each momentum integral we have a factor $(\alpha_s \ln p^2 \ln 1/x)$.
It is this kind interaction between the two ladders which we will study in the 
following, staying within the double logarithmic approximation of gluon ladders.     
Particular attention will be given to the possibility that this
recombination of two chains takes place in the perturbative region, i.e. 
in the region of large transverse momenta. 

Recently 
an important potential application of such recombination effects 
has been suggested. In their attempt to explain the ridge effect reported  
by the CMS group at the LHC \cite{Khachatryan:2010gv}, it has been suggested \cite{Dumitru:2010iy} 
that the observed long range rapidity correlation and azimuthal correlation 
can be explained, within the Color Glass Condensate framework, by a two-chain 
recombination which will be discussed in this paper.   

\section{Two noninteracting ladders} 
We begin with the double logarithmic 
approximation of the two-chain configurations shown in Figs.1 and 2: we search 
for regions of integration where each closed momentum loop gives two logarithms, 
one in the transverse momentum and one in rapidity. We restrict ourselves to 
gluon ladders which, at small $x$, are known to give the largest contributions.
We parametrize our momenta as
\beq
k = x p_A + y p_B + \bk
\label{eq:Sudakov}    
\eeq
where $p_A$, $p_B$ are the large momenta of the incoming protons $A$ and $B$, resp., 
the momentum fractions $x$, $y$ range between $0$ and $1$, and $\bk$ denotes 
the two dimensional transverse momentum.  
In the double logarithmic approximation, the BFKL kernel and the splitting 
function $P_{gg}$ lead to the same answer (Fig.3):
we can either start from the small-$x$ limit which is described by the 
BFKL equation and then take the limit of strongly different momentum scales;
alternatively, we can begin with the collinear limit where the DGLAP equations 
apply and then take the limit of small $x$.   
For our purposes it will turn out that the approach based upon the 
BFKL equation is more suitable: it is the region where the logarithms in $1/x$ 
are slightly larger than the transverse momentum logarithms where the recombination 
effects become important.   
\begin{figure}[t]
\begin{center}
\epsfig{file=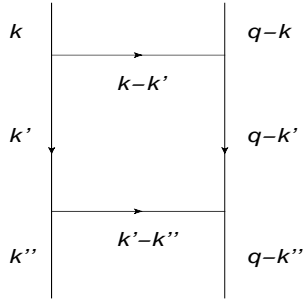,width=4cm,height=4cm}\\
\caption{A single cell inside a ladder, below the produced pair of jets.} 
\end{center}
\end{figure}

In the region of strongly ordered transverse momenta the color singlet BFKL 
kernel is approximated by 
\beq       
g^2 \left(-\bq^2 + \frac{\bk^2 (\bq-\bk)^2 + {\bk'}^2 (\bq-\bk')^2}{(\bk-\bk')^2}\right)
\approx 2 g^2 \bk' (\bk'-\bq)
\label{eq:DLABFKL}
\eeq
This approximation will be used in the following.

We begin with the two noninteracting ladders 
shown in Fig.1. Our main focus 
is on the integration over the loop momentum $\bq$, and we consider a single 
cell with momentum $\bk'$ inside one of the ladders below the produced pairs of jets. 
This cell is illustrated in Fig.3.
Using (\ref{eq:DLABFKL}) for the upper rung 
(and the corresponding expression for the lower rung), together with the 
gluon propagators for the vertical lines, one sees that the 
integration over the transverse momentum $\bk'$ 
is logarithmic only if $k'  \gg q$, i.e. $q$ defines the momentum scale 
$Q_0$ where the $\bk^2$ evolution along the ladders starts. 
On the other hand, the range of the integration over the 
momentum transfer along the ladders is determined by the size of the interaction region 
and by the correlation length of the initial gluons 
of the two ladders inside the proton: we denote this effective radius by $\tilde R$, and 
put $Q_0^2 = 1/{\tilde R}^2$.
As a result, the cross section for the production of two pairs of gluon jets 
(cf. Fig.1) will be of the form:
\beq
\frac{d \sigma}{dY_1 dY_2 d^2 \bp_1 d^2 \bp_2} \sim
\frac{1}{{\tilde R}^2} \frac{1}{(\bp_1^2)^2} \frac{1}{(\bp_2^2)^2} f(x_1,\bp_1^2) f(y_1,\bp_1^2) 
              f(x_2,\bp_2^2) f(y_2,\bp_2^2)
\label{eq:crosssection1}
\eeq
where $f(x,\bp^2)=xg(x,\bp^2)$ denotes the gluon density with initial 
momentum scale $Q_0^2$, the factor 
$1/{\tilde R}^2$ results from the integration over the loop momentum $\bq$,
and the momentum factors $1/(\bp_1^2 \bp_2^2)^2$ 
represent the two production vertices of the pairs of gluon jets 
(evaluated in the double logarithmic approximation).   

As an important feature of (\ref{eq:crosssection1}) we mention that, 
as long as interactions between different parton chains are not taken 
into account, the momentum transfer along a ladder is of the order of the 
initial momentum scale of the (multi)parton distribution.       

\section{Recombinations in a two-ladder configuration}
We now turn to the main topic of this paper, 
a study of the two recombinations shown in Fig.2. 
We begin with the recombination 
vertex below the produced jets; the kinematics are illustrated in Fig.4,.
Above the recombination, the two ladders formed by the lines (13) and (24) 
are in the color singlet configuration; below the color singlet ladders 
are formed out of (14) and (23). As a result, each color 
recombination is accompanied by the color suppression factor 
\beq
\frac{1}{N_c^2 -1}
\label{eq:colorsuppression}
\eeq   
\begin{figure}[t]
\begin{center}
\epsfig{file=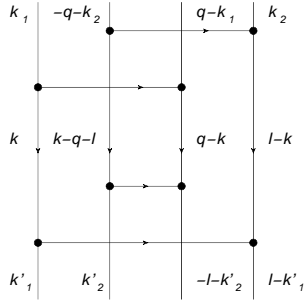,width=4cm,height=4cm}\\
\caption{Kinematics of a recombination $(14)(23) \to (13)(24)$.}
\end{center} 
\end{figure} 
In order to have, for the momentum loops below the lower two rungs (connecting 
$(14)$ and $(23)$), logarithmic momentum integrals, we need 
\beq
|\bk'_1|, |\bk'_2| \gg |\bl|. 
\label{eq:restr1}
\eeq
Similarly, in order to find logarithmic momentum integrals above 
the upper two rungs (connecting (13) and (24)), we need   
\beq
|\bk_1|, |\bk_2| \gg |\bq|,
\label{eq:restr2}
\eeq
Finally, inside the $\bk$ loop of Fig.4. we need 
\beq
|\bk_1|, |\bk_2| \gg |\bk| \gg |\bk'_1|, |\bk'_2|,\,\,\, |\bk| \gg |\bl|,\,\,\,
|\bk_1|, |\bk_2| \gg |\bk'_1|, |\bk'_2|.
\label{eq:restr3}
\eeq
Using, for the upper two rungs, the approximations following from (\ref{eq:DLABFKL}), and 
combining them with the propagators for the vertical gluon lines, we obtain:
\beq
(2 N_cg^2)^2 \int \frac{d^2 k}{(2\pi)^3}  \int \frac{d^2 q}{(2\pi)^3} \frac{2 \bk(\bk-\bq)}{\bk^2 (\bq-\bk)^2}  \frac{2 \bk(\bk-\bq)}{\bk^2 (\bq-\bk)^2}     
\label{eq:recvertex1}
\eeq
In order to obtain a logarithmic integrals of $\bk$, we identify two regions of $\bk$ and $\bq$: 
either $|\bq| \gg |\bk$ or  $|\bq| \gg |\bq - \bk|$. For each of the two cases, after averaging 
over the azimuthal angle, we arrive at the integrals
\beq  
2 g^4 \int \frac{d^2 q}{(2\pi)^3} \frac{1}{\bq^2} \int^{q^2} \frac{d^2 k}{(2\pi)^3} \frac{1}{\bk^2}
\label{eq:recvertex2}
\eeq
which gives the desired logarithmic integral in $\bk$ (or ($\bq-\bk$)). 

For the two ladders below the recombination we derive, from the condition 
(\ref{eq:restr3}) and from the second integral in (\ref{eq:recvertex2}), 
that the upper cutoff is given by $q^2$. 
The lower cutoff is obtained by applying the 
discussion of section 2.1: the loop momentum $\bl$ appears only 
inside the initial condition of the 
lower proton $B$, and it is restricted by the effective scale $Q_0^2$
The condition (\ref{eq:restr2}) implies that the momentum $\bq$ also defines 
the lower momentum cutoff for the ladders above the recombination vertex. 

In order to find the full dependence on $q^2$, we need to consider the 
full diagram in Fig.2 \footnote{Otherwise the momentum $\bq$ will run through the blob corresponding to the proton initial conditions and, like the momentum $\bl$, it will be restricted by a low scale $Q_0$.}. The recombination vertex above the produced pairs of 
jets is analysed in the same way as the lower one. This leads to an expression 
similar to (\ref{eq:recvertex2}), i.e. the complete dependence in $\bq$ is of the form:
\beq
\int \frac{d^2 q}{(\bq^2)^2},
\label{eq:qdependence}
\eeq 
The integral in $\bq$ is dominated by small values. Since the momentum $\bq$ 
defines the upper momentum cutoff, both for the two ladders below the lower 
recombination vertex and for the two ladders above the upper recombination 
vertex, we conclude that the infrared divergence of the $\bq$-integration 
destroys the ladders above and below the recombination vertices. 
The recombination vertices, therefore, are absorbed into the 
nonperturbative initial conditions. 
As far as the perturbative part is concerned, we are back to the two 
noninteracting chains of section 2.1. 

The situation changes if logarithms in rapidity become more important than those in transverse momentum,
i.e. within the BFKL approach we move towards small $x$. In Fig.2, we replace the rungs by BFKL Green's 
functions. Re-drawing the diagram in an more suitable way, we arrive at Fig.5: 
\begin{figure}[t]
\begin{center}
\epsfig{file=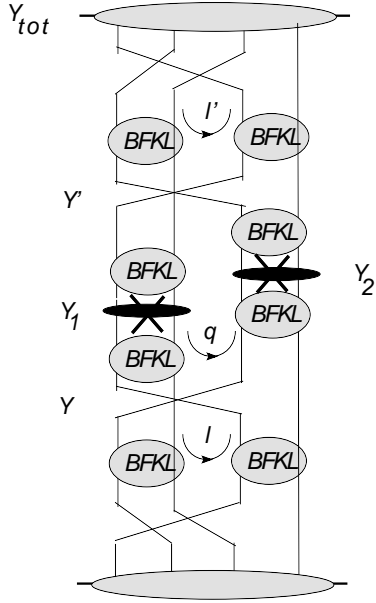,width=5cm,height=8cm}\\
\caption{Another way of drawing Fig.2b}
\end{center}  
\end{figure}
Let us first reformulate the result which we have just obtained 
in the double logarithmic approxmation. 
We have shown that, in order to find inside the BFKL Green's function the maximal number of 
transverse logarithms, the momentm transfer across the Green's function has to be smaller than 
the transverse momenta of the two gluons entering the Green' function at the low-momentum 
side. In Fig.5. this says that $\bl$ and $\bl'$ have to be small, while for the 
$q$-loop we found the integral $\int dq^2/q^4$ which favors small values, too. 

In order 
to see how the appearance of large rapidity intervals changes this situation, let us use 
the following integral representation for the forward BFKL-Green's function:
\beq
G(k,k';Y-Y')=\int \frac{d \omega}{2\pi i} e^{\omega(Y-Y ')} \int \frac{d \nu}{2\pi i}
\left( \frac{k^2}{{k'}^2} \right)^{\mu} \frac{1}{\omega -\chi(\mu,0)}, 
\label{Greensfunction} 
\eeq 
where $\mu =  i \nu + \frac{1}{2}$, the integration contours in $\mu$ and $\omega$ run 
parallel to the imaginay axis, $\chi(\mu,n)$ is the BFKL eigenvalue function, and we 
have averaged over the azimuthal 
angles of $\bk$, $\bk'$. Furthermore, we have kept only the leading term of the conformal 
spin, $n=0$. From this representation one easily deduces the dependence upon 
$Y-Y'$ and $\ln \frac{k^2}{{k'}^2}$: after the integration over $\omega$ the 
saddle point analysis of the remaining $\mu$ -integral shows that, 
for large $\ln \frac{k^2}{{k'}^2}$, the dominant contributions come from 
$\mu = i \nu + \frac{1}{2} \approx 0$ and $\omega={\cal O}(1)$, whereas 
for large $Y-Y'$, one finds $\mu \approx \frac{1}{2}$ and 
small $\omega = \omega_{BFKL}= \frac{4 N_c \ln2 \alpha_s}{\pi}$. 
This observation has important 
consequences. 
Let us denote the rapidities of the two recombinations by $Y'$ and $Y$, resp..
Beginning with the BFKL amplitudes near the produced jets, large 
rapidity intervals $Y'-Y_1$, $Y_1 -Y$, $Y'-Y_2$, $Y_2-Y$ increase the 
corresponding anomalous dimensions and hence favor larger values of the 
central loop momentum, $\bq$. To see this im more detail, we    
insert the integral representations for all the BFKL Green' function in
Fig.5. We arrive at                
$$
\frac{d \sigma}{dY_1 dY_2 d^2 \bp_1 d^2 \bp_2} \sim
%\frac{1}{{\tilde R}^2}
 \frac 1{R^2_c}\frac{1}{R_c^2}
\frac{1}{(\bp_1^2)^2} \frac{1}{(\bp_2^2)^2}
\int \frac{d \mu'}{2 \pi i} \int \frac{d \mu}{2 \pi i} 
\int \frac{d \mu'_1}{2 \pi i} \int \frac{d \mu_1}{2 \pi i}
\int \frac{d \mu'_2}{2 \pi i}\int \frac{d \mu_2}{2 \pi i}\cdot $$
$$ \int dY \int dY' \int \frac{d^2 q}{q^4} 
\Big[ \left( \frac{\bq^2}{Q_0^2} \right)^{\mu'} e^{(Y_{tot}-Y') \chi(\mu')} 
\Big]^2 \cdot
$$
\beq
\Big[ \left( \frac{\bp_1^2}{\bq^2} \right)^{\mu_1'} e^{(Y'-Y_1) \chi(\mu_1')} \Big]
\Big[ \left( \frac{\bp_1^2}{\bq^2} \right)^{\mu_1} e^{(Y_1-Y) \chi(\mu_1)} \Big]
\Big[ \left( \frac{\bp_2^2}{\bq^2} \right)^{\mu_2'} e^{(Y'-Y_2) \chi(\mu_2')} \Big]
\Big[ \left( \frac{\bp_2^2}{\bq^2} \right)^{\mu_2} e^{(Y_2-Y) \chi(\mu_2)} \Big]\cdot
\nonumber\eeq
\beq
\cdot
\Big[ \left( \frac{\bq^2}{Q_0^2} \right)^{\mu} e^{Y \chi(\mu)} \Big]^2 
\label{eq:BFKLcrosssection} 
\eeq
Here we have introduced another length scale, $R_c^2$: it results from the 
additional integrals (as compared to eq.(\ref{eq:crosssection1})) $d^2\bl$ and $d^2\bl'$, 
which are restricted by the proton radius and by the correlation between the 
two chains inside the proton.  
As long as all BFKL amplitudes are in DGLAP regime, i.e. they are dominated 
by the logarithms in the transverse momenta, all $\mu$ variables are small, 
and we are in the situation which we have described above: 
the $\bq$ integral is dominated 
by small values, and the transverse momentum  
logarithms inside those four BFKL amplitudes which are close to the protons 
are destroyed.  If, however, the rapidity intervals become large 
and the BFKL amplitudes are in the 
small-$x$ region, $\mu$ values are close to   
$\frac{1}{2}$. As a result, in (\ref{eq:BFKLcrosssection}) the overall power 
of $\bq^2$ may increase and the dominance of the small-$\bf q^2$ region 
disappears. One easily sees that large rapidity intervals 
near the protons, $Y_{tot}-Y'$ and $Y$, tend to make $\mu$ and $\mu'$ large and
thus help to increase the overall power of $q^2$.      

Let us see in more detail how this balance works. 
Defining in (\ref{eq:BFKLcrosssection}) the phase function
$$
\Phi(\mu����,\mu'_1,\mu_1,\mu'_2,\mu_2,\mu)=
2\left( (Y_{tot}-Y')\chi(\mu') + \mu' \ln \frac{q^2}{Q_0^2} \right)+$$
\beqn
+\left( (Y'-Y_1)\chi(\mu'_1)+ \mu'_1 \ln \frac{p_1^2}{q^2}\right)+
\left((Y_1-Y)\chi(\mu_1)+ \mu_1 \ln \frac{p_1^2}{q^2}\right)
\nonumber\\
+\left((Y'-Y_2)\chi(\mu'_2)+ \mu'_1 \ln \frac{p_2^2}{q^2}\right)+
\left((Y_2-Y)\chi(\mu_2)+ \mu_1 \ln \frac{p_2^2}{q^2}\right)
\nonumber\\
+2\left(Y \chi(\mu) + \mu\ln \frac{q^2}{Q_0^2} \right),
\label{eq:saddle}
\eeqn
the saddle points are determined from the conditions:
\beq
0 = \frac{\partial \Phi}{\partial \mu} = 2\chi'(\mu_s) Y + 
2\ln  \frac{q^2}{Q_0^2},
\label{eq:nearprotoncond}
\eeq
and 
\beq
0 = \frac{\partial \Phi}{\partial \mu_1} = \chi'(\mu_{1,s}) (Y_1-Y) + 
\ln  \frac{p_1^2}{q^2},
\label{nearvertexcond}
\eeq
which lead to  
\beq
 \chi'(\mu_s) = - \frac{\ln  \frac{q^2}{Q_0^2}}{Y}
\label{eq:nearproton}
\eeq   
and 
\beq
 \chi'(\mu_{1,s}) = - \frac{\ln  \frac{p_1^2}{q^2}}{Y_1-Y}
\label{eq:nearvertex}
 \eeq
Similar equations are obtained for the other $\mu$ variables.

For a systematic analysis one first determines, for fixed $Y$, $Y'$ and 
$q^2$, the stationary points of the $\mu$ variables and then finds 
the dominant values of the rapidities $Y$, $Y'$ and of the momentum 
scale $\ln q^2$.
As we have said before, if in (\ref{eq:nearproton}) the evolution in 
rapidity dominates over that in momentum scale, the rhs becomes small. 
Since $\chi'(\mu)$ vanishes at $\mu=\frac{1}{2}$, we have  
\beq
\mu_s \approx \frac{1}{2} - \frac{1}{\chi''(\frac{1}{2})} 
\frac{\ln \frac{q^2}{Q_0^2}}{Y}.
\label{eq:saddlepoint1}
\eeq
On the other hand, if in (\ref{eq:nearvertex}) the interval in momentum 
evolution is larger than in rapidity, the rhs is large when $\mu$ is close 
to zero:  
\beq
\mu_{1,s}\approx \sqrt{a \frac{ Y_1-Y}{\ln  \frac{p_1^2}{q^2}}}
\label{eq:saddlepoint2}
\eeq
with $a=\frac{N_c\alpha_s}{\pi}$ and $\chi''(\frac{1}{2})=a28\zeta(3)$
(here $\zeta(3)\approx 1.202$ denotes the Riemann zeta function). 

The results can be illustrated in terms of evolution paths in Fig.6.
There is an infinite number of paths in the $\ln k^2$-$y$ plane 
which connect the protons with the produced jets 
with rapidity $Y_1$, $Y_2$  and momenta  $\bp_1^2$, $\bp_2^2$.
The saddle point analysis determines the most probable path.
Two examples are shown in Fig.6.   
\begin{figure}[t]
\begin{center}
\epsfig{file=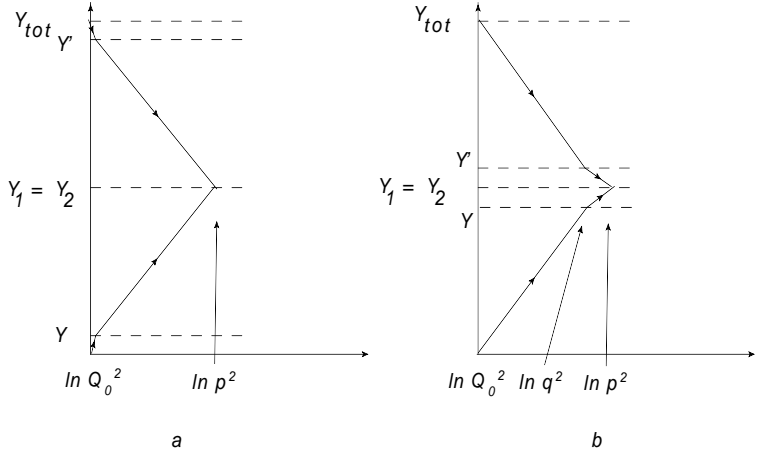,width=10cm,height=6cm}\\
\caption{Two different paths of evolution: (a) Normal path which disfavors 
recombination, (b) a path which supports a recombination.    
For simplicity, we have taken $\bp_1^2=\bp_2^2= p^2$ and $Y_1 = Y_2$.} 
\end{center} 
\end{figure}
Let us now consider a few special cases. For simplicity, we start with the 
symmetric choice $Y_1=Y_2$ and $p_1^2=p_2^2=p^2$. In order to get a large $q^2$ we search for the 
situation where $\mu>\mu_i$ and $\mu'>\mu'_i$. 
We insert the saddle point values (\ref{eq:saddlepoint1}), 
(\ref{eq:saddlepoint2}) into (\ref{eq:saddle}), and first look 
for the extrema with respect to $Y$ and $Y'$, that is for the saddle 
point of the expression
\begin{equation}
A_{lower}=e^{\sqrt{4a(Y_1-Y)\ln(p^2_1/q^2)}} e^{2Y\chi(\mu_s)}
e^{\sqrt{4a(Y_2-Y)\ln(p^2_2/q^2)}}
\label{eq:alow}
\end{equation} 
which belongs to the lower part of Fig.5.
To get (\ref{eq:alow}) we have used the value of 
$\mu_s$ in (\ref{eq:saddlepoint1}) 
and neglected a weak dependence of 
$\mu_s\sim 1/2$ on $Y$ coming from (\ref{eq:saddlepoint1}). 
Thus the typical value of $Y$ is
\beq
Y=Y_s=Y_1-a\frac{\ln^2(p^2/q^2)}{\chi^2(\mu_s)},
\label{deltay1}
\eeq
leading to
\beq
A_{lower}\sim \exp\left(\frac {2a}{\chi(\mu_s)}\left[2\ln^{3/2}(p^2/q^2)
-\ln^2(p^2/q^2)\right]+2\chi(\mu_s)Y_1\right)\, .
\label{eq:slow}
\eeq
An analogous expression is obtained for the upper part of the amplitude in 
Fig.5, $A_{up}$.
Assuming that the total avaliable rapidity interval $Y_{tot}$ 
and the sub-rapidities $Y_1,\ Y_2$ are so large that the saddle point 
position $\mu_s$ (\ref{eq:saddlepoint1}) is close to 1/2, 
i.e. $28\zeta(3)aY_1>>\ln(q^2/Q^2_0)$. We put $\mu_s=0.5$ in (\ref{eq:alow}), 
(\ref{eq:slow}), and we see that in (\ref{eq:BFKLcrosssection}) the integral 
over $q^2$ takes the form
\beq
\int \frac{d\ln q^2}{q^2}\exp\left(\frac 1{\ln 2} 
\left[2\ln^{3/2}(p^2/q^2)-\ln^2(p^2/q^2)\right]\right).
\label{eq:qsaddle}
\eeq
Here we have used the LO BFKL ratio $a/\chi(\mu_s)=a/\chi(0.5)=1/4\ln 2$.
The integral (\ref{eq:qsaddle}) has 
its saddle point at
\beq
q^2\sim p^2\exp(-z^2)
\label{qsad}
\eeq
with $z=3/4+\sqrt{9/16+0.5\ln 2}\simeq 1.7$; $\exp(-z^2)\simeq 0.055$.
The corresponding value of $\mu_{1,s}\simeq z/4\ln 2\sim 0.6$ is not  small enough to justify the approximate estimate (\ref{eq:saddlepoint2}).
We therefore conclude that the competition between the $Y$ 
and $q^2$ dependence leads to a rapidity saddle point (\ref{deltay1}) 
somewhere inside the avaliable interval $(Y_1,0)$, which in its turn leads to a 
rather large $\mu_{i,s}$ violating the initial inequality $\mu_i<\mu$.

In general we can say that the ordering $\mu_{i,s} < \mu_s$ leads to 
the opposite ordering of the intercepts, $\chi(\mu_s) < \chi(\mu_{i,s})$.
Therefore, in (\ref{eq:BFKLcrosssection}) the dominant contribution to the 
$Y$ integral comes from the region of small $Y$ where the recombination vertex 
is close to the initial proton and far from the production vertex of the 
dijets. For a small $Y$ value the anomalous dimension $\mu$ cannot be large,
and the essential $q^2$ values are small as well. This means that we are back 
to the situation of non-interacting ladders, illustrated in Fig.6a.    

Next we turn to the opposite case  $\mu_i>\mu$ which corresponds 
to $\chi(\mu_i)<\chi(\mu)$. Now the dominant $Y$-value is large and close 
to the rapidity of high $E_T$ dijets $Y_1$. However, now the whole anomalous 
dimension in the $q^2$ behaviour $2(\mu+\mu')-\mu_1-\mu'_1-\mu_2-\mu'_2<0$ is 
negative, and the $q^2$ integral is dominated by a low $q^2$-value.\\

The most interesting possibility is to put the recombination vertices just 
as close as possible to the high $E_T$ dijet production matrix elements. 
In this case there is no BFKL or DGLAP evolution in the intervals between 
the produced pairs of jets and the recombination vertices. 
That is, in the centre of Fig.5,  we simply delete the four 'BFKL blobs' 
nearest to the produced jet pairs. Correspondingly, 
in (\ref{eq:BFKLcrosssection}) we eliminate the third line, togather 
with the integrations over
$\mu_1,\mu_2,\mu'_1,\mu'_2$. The rapidities $Y,Y'$ are close to $Y_i$, 
and the $q^2$ integral takes the form
\beq
\int\frac{d\ln q^2}{q^2}q^{4(\mu_s+\mu'_s)} \ ,
\label{q-int}
\eeq
where the saddle point values, $\mu_s$ and $\mu'_s$, 
follow from the condition (\ref{eq:nearprotoncond}):  
\beq
0 = \chi'(\mu_{s}) Y + 
\ln  \frac{q^2}{Q_0^2}
\eeq
and 
\beq
0 = \chi'(\mu'_s) (Y_{tot}-Y') + 
\ln  \frac{q^2}{Q_0^2}\,.
\eeq
Their values are taken from (\ref{eq:saddlepoint1}):
\beq
\mu_s \approx \frac{1}{2} - \frac{1}{\chi''(\frac{1}{2})} 
\frac{\ln \frac{q^2}{Q_0^2}}{Y}\,,
\eeq
i.e. the integral over $q^2$ receives its main contribution from $q^2$ close to $\min\{p_1^2,p_2^2\}$~~\footnote{In the region of $q>p_i$ the momentum $q$ will destroy the matrix element of high
$E_{T,i}$ dijet production replacing in (\ref{eq:BFKLcrosssection}) the factor $1/p^4_i$ by $1/q^4$.}
This situation belongs to the evolution path shown in Fig.6b. 

Let us finally consider a more realistic situation with $Y_1=Y_2$ but $p_2 < p_1$. Recall that the 
true argument of the BFKL amplitude is not rapidity
but the momentum fraction $x$, that is actually we have to take $Y=\ln(1/x)$. When $p_1>>p_2$  
for the same rapidities $Y_1=Y_2$ we get in the right ladder the momentum fraction $x_2<<x_1$. 
In other words, in this configuration we may put, in Fig.5, the recombination vertex just into the cell 
nearest to the left dijet. But then there will be a large $\ln x$ (and may be $\ln q^2$) 
interval for the evolution of the right ladders (between the dijets on the rhs and the two recombination vertices. 
In other words in Fig.5. we delete only the two 'BFKL blobs' on the lhs 
below and above the dijet production.
Assuming that, in (\ref{eq:BFKLcrosssection}), the total rapidity interval $Y_{tot}$ is very large, we may perform first the rapidity integral
\beq
\int dY\exp[-Y(\chi(\mu_2)-2\chi(\mu))]=\frac 1{\chi(\mu_2)-2\chi(\mu)}
\eeq
where for the BFKL blobs on the lhs we have set $\chi(\mu_1)=0$, and for $\mu$ we put its asymptotic value $\mu=1/2$.
Now we close the countor of the $\mu_2$ integration around the pole $\chi(\mu_2)-2\chi(\mu)=0$: this leads to $\mu_2\simeq 0.18$. The same result is 
obtained for $\mu'_2$.
Finally, the $q^2$ integral takes the form
\beq
\int^{p^2_2} d\ln q^2 q^{2(1-\mu_2-\mu'_2)},
\eeq 
and the major contribution comes from the domain close to upper limit $q^2\sim p^2_2$.\\

A closer look reveals still another detail. In the region of interest, for example in a 14 TeV $pp$-collision at the LHC,  we observe in the central region 
the dijet with  $p_1\sim 20 GeV$, corresponding to $x\sim 2p_1/\sqrt{s}\sim 0.003$. For such $x$-values, the anomalous dimension observed at HERA is not so large. 
For $x<0.01$ the behaviour of the structure function $F_2(x,q^2)$ can be parametrized as
\beq
F_2= c(q^2)x^\lambda
\label{f2}
\eeq 
with $c\simeq const(q^2)$ and $\lambda=0.048\pm0.004$~\cite{h1}.
This means that effective anomalous dimension $\mu_{eff}=\lambda\ln(1/x)\sim 0.28$ for $x=3\cdot 10^{-3}$.
This value is still large enough to provide the convergence of the $q^2$ integral (\ref{q-int}) in the large $q^2$ domain 
for the case considered above where both recombination vertices are jsut near the dijet production cell. 
However it is not evident that the parametrization (\ref{f2}) reflects the behaviour of a {\it single} ladder. At not large $q^2$
the experimentally measured $F_2$ already includes some 
absorptive effects which reduce the growth of $F_2$ with $x$ decreasing and thus 
leads to a lower value of $\lambda$ in comparison with a single ladder contribution. In other words the true value of $\mu_{eff}$ which corresponds to a single ladder may be even larger, pushing the
characteristic values of $q^2$ closer to the (lower) hard scale $p^2_2$.\\

\section{Generalizations}

So far we have discussed the effect of two recombinations inside a two-chain contribution:
one recombination on esch side of the produced jet pairs. 
Let us first comment on the case where we have no second recombination vertex 
above the jet pairs: as far as 
%the 
 only one 
recombination vertex is concerned, the integration over $\bq$ 
is logarithmic. However, $\bq$ runs also through both upper ladders and defines  
the low momentum scale $Q_0^2$ where the evolution starts: a large value of $\bq$ therefore kills 
the evolution in the upper ladders, whereas a low value prevents the evolution in the lower ladders.
Therefore, a single recombination vertex is suppressed.

Next a comment on the color suppression factor (\ref{eq:colorsuppression}). This suppression 
applies to the case when, as illustrated in Fig.2, there is evolution above and below the recombination vertex.
As we have discussed before, in a preferred situation we have little or no evolution between the 
recombination vertices and the dijet production vertices. In this case there is no need to reconnect, between the two 
recombination vertices, the four t-channel gluon lines to color singlet pairs.  
As result, the color suppression becomes much weaker..    

Next we consider the case of more than two chains, say three chains with three 
produced pairs of jets..
In this case a pair of two recombination vertices can be attributed to any pair of chains, i.e. 
we have three possibilities. Similarly, for $n$ chains we have 
$\frac{n(n-1)}{2}$ possiblities: these counting factors can easily overcome 
the color suppression factor in (\ref{eq:colorsuppression}). As an example, for 
$n=4$, the overall counting factor is already $3/4$, and it exceeds unity
for $n\ge5$.

Finally, we mention another important possibility, related to final states 
with rapidity gaps.
Besides the recombination illustrated in Fig.2 there exists another 
configuration to which our discussion applies. We show this in Fig.7:
\begin{figure}[t]
\begin{center}
\epsfig{file=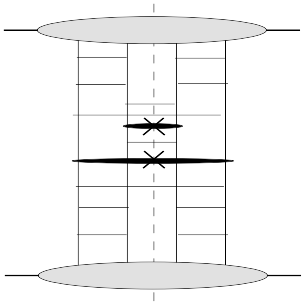,width=4cm,height=4cm}\\ 
\caption{A recombination of two ladders which allows for diffractive 
states.} 
\end{center}
\end{figure}  
Applying our previous discussion, in particular the evolution paths 
illustrated in Fig.6, we conclude that the momentum scale at the upper end 
of the lower rapidity gap, $q^2$, will be above $Q_0^2$ but not too close 
to the jet momenta $p_1^2=p_2^2$: this allows for 'semihard' diffraction and 
is in qualitative agreement with inclusive diffraction seen at HERA.    

\section{Conclusions and outlook}
We have studied the possibility of interactions ('recombinations') between two evolution chains, 
which describe double parton scattering corrections in high energy hadron collisions.
We show that, thanks to large anomalous dimensions of the parton 
distributions in the low-$x$ region, such an interaction may occur 
at small distances within the pertubative domain, provided we consider {\it two} 
recombination vertices (which describe the chain-chain 
interaction) placed relatively close to the 'hard' matrix elements.

In our double leading log analysis of gluon ladder diagrams we find a 
competition between 'collinear' logarithms ($\ln \bp^2$) and 'energy' logarithms ($\ln(1/x)$).
Depending on the ratio between these logarithms, the major contribution to integrals over the 
rapidity of the two recombination vertices, $Y$ and $Y'$, comes either from the region near the protons (Fig.6a) 
or from the region close to the jet production vertices (Fig.6b), i.e. these integrals have no saddle point somewhere in centre 
of the avaliable interval. The first case (Fig.6a) corresponds to two independent ladders which do not communicate with each other and 
are described by 'double DGLAP' evolution equations. More interesting is the second possibility (Fig.6b) where the recombination 
vertices are close to the hard matrix elements and thus are entirely in the perturbative region.
These configurations may lead to nontrivial correlations between the secondaries produced in 'double parton scattering' processes. 
We note that, in Fig.6b,  the rapidities and transverse momenta of the partons inside the 
recombination vertices need not be very close to the 'hard' matrix element: in a more or less realistic situation the convergence of the 
integrals in rapidity ($Y$ and $Y'$) and transverse momentm ($q$) are rather slow, since they are driven by numerically small powers
of $1/x$ and $q^2$. Therefore the particles coming from the recombination vertices may still be separated from those  
produced via the 'hard' subprocess by relatively large intervals in rapidity ($\sim$ few units) and in the logarithms of transverse 
momenta.

The interactions between different ladders discussed in this paper also allow for semihard diffractive final states (Fig.7). 

In the case of multiple parton interactions with a larger number of evolution chains the suppression of chain-chain 
interactions caused by the colour factor $1/(N^2_c-1)$ may be compensated by the 
combinatorical factor. For $n$ chains we have $n(n-1)/2$ possibilities. According to naive 'eikonal model' estimates the mean
number of chains in proton-proton collision at the LHC is 
$<n>\sim 5$. Therefore the expected probability of multi-chain recombinations is not small and may lead both to a noticeable correlation 
between the secondaries in inclusive processes and to 'semihard' diffraction final states.

All results of this paper are based upon the double logarithmic 
approximation (with fixed $\alpha_s$). We consider this as a first step towards a more accurate 
analysis. Within the small-$x$ approach it is possible to go beyond the 
double logarithmic approximation and to reach single logarithmic accuracy 
(leading $\ln 1/x$). Also, a more detailed numerical analysis will be needed in order 
to obtain a more reliable estimate of the importance of the recombination 
corrections addressed in this paper. Both tasks will be topics of future work.  

We finally mention an important consequence of our result. In contrast to 
non-interacting multiparton chains which often are modelled within the 
eikonal approximation, corrections due to the recombination of ladder 
diagrams no longer fit into the eikonal picture. This raises the question 
of the AGK cutting rules which provide a crucial theoretical constraint 
of multiparton corrections. An investigation of this problem is quite 
important. \\
\\ 
{\bf Acknowledgements:}\\ 
The work by MGR was supported by the grant RFBR 11-02-00120a and by the Federeal Program of the Russian State RSGSS-65751.2010.2.

\end{document}